\documentclass{aa}  

\usepackage{graphicx}
\begin{document} 


\newcommand{\amu}{\ensuremath{\mathrm{m}_{\rm amu}}}                 
\newcommand{\msunyr}{\ensuremath{\mathit{M}_{\odot}{\rm yr}^{-1}}}   
\newcommand{\kms}{\ensuremath{{\rm km\,s^{-1}}}}                   
\newcommand{\msun}{\ensuremath{\mathit{M}_{\odot}}}   
\newcommand{\mini}{\ensuremath{M_{\rm ini}}}                         
\newcommand{\msunano}{\ensuremath{\mathit{M}_{\odot}{\rm yr}^{-1}}}   
\newcommand{\lsun}{\ensuremath{\mathit{L}_{\odot}}}                  
\newcommand{\rsun}{\ensuremath{\mathit{R}_{\odot}}}                  
\newcommand{\zsun}{\ensuremath{\mathit{Z}_{\odot}}}                  
\newcommand{\tin}{\ensuremath{\mathit{T}_{\mathrm{in}}}}

\newcommand{\gstar}{\ensuremath{\mathrm{g}_{\star}}}                 
\newcommand{\geff}{\ensuremath{\mathrm{g}_{\rm eff}}}                
\newcommand{\logg}{\ensuremath{\log \mathrm{g}}}                     
\newcommand{\lstar}{\ensuremath{\mathit{L}_{\star}}}                 
\newcommand{\mdot}{\ensuremath{\dot{M}}}                             
\newcommand{\mstar}{\ensuremath{\mathit{M}_{\star}}}                 
\newcommand{\rstar}{\ensuremath{\mathit{R}_{\star}}}                 
\newcommand{\teff}{\ensuremath{\mathit{T}_{\rm eff}}}                
\newcommand{\reff}{\ensuremath{\mathit{R}_{\rm phot}}}                
\newcommand{\vinf}{\ensuremath{\upsilon_{\infty}}}                          
\newcommand{\vesc}{\ensuremath{\upsilon_{\rm esc}}}                         
\newcommand{\tstar}{\ensuremath{\mathit{T}_{\star}}}                 
\newcommand{\K}{\ensuremath{\mathrm{K}}}                 

\newcommand{\vrot}{\ensuremath{\upsilon_{\rm rot}}}                         
\newcommand{\vcrit}{\ensuremath{\upsilon_{\rm crit}}}                         
\newcommand{\vphot}{\ensuremath{v_{\rm phot}}}                         
\newcommand{\vblack}{\ensuremath{v_{\rm black}}}                         
\newcommand{\vedge}{\ensuremath{v_{\rm edge}}}                         
\newcommand{\vwind}{\ensuremath{v_{\rm wind}}} 

\newcommand{\vs}{\ensuremath{v_{\rm s}}}                         
\newcommand{\taulyc}{\ensuremath{\tau_{\mathrm{Lyc}}}}                 
\newcommand{\tflow}{\ensuremath{\tau_{\mathrm{flow}}}}                 
\newcommand{\trec}{\ensuremath{\tau_{\mathrm{rec}}}}                 
\newcommand{\tvar}{\ensuremath{\tau_{\mathrm{var}}}}                 
\newcommand{\tsurf}{\ensuremath{T_{\mathrm{surf}}}}                 
\newcommand{\teffg}{\ensuremath{T_{\mathrm{eff,GVA}}}}                 
\newcommand{\teffgcorr}{\ensuremath{T_{\mathrm{eff,GVAcorr}}}}                 
\newcommand{\teffc}{\ensuremath{T_{\mathrm{eff,CMF}}}}                 

\newcommand{\xsurf}{\ensuremath{X_{\mathrm{sur}}}}                 
\newcommand{\ysurf}{\ensuremath{Y_{\mathrm{sur}}}}                 
\newcommand{\csurf}{\ensuremath{C_{\mathrm{sur}}}}                 
\newcommand{\nsurf}{\ensuremath{N_{\mathrm{sur}}}}                 
\newcommand{\osurf}{\ensuremath{O_{\mathrm{sur}}}}                 
\newcommand{\lam}{\ensuremath{\lambda}}                 

\newcommand{\magu}{\ensuremath{\mathit{M}_{\rm U}}}              
\newcommand{\magb}{\ensuremath{\mathit{M}_{\rm B}}}           
\newcommand{\magv}{\ensuremath{\mathit{M}_{\rm V}}}    
\newcommand{\magr}{\ensuremath{\mathit{M}_{\rm R}}}    
\newcommand{\magi}{\ensuremath{\mathit{M}_{\rm I}}}    
\newcommand{\magk}{\ensuremath{\mathit{M}_{\rm K}}}    

\newcommand{\mbol}{\ensuremath{\mathit{M}_{\rm bol}}}   
\newcommand{\ang}{\ensuremath{\text{\AA}}}                  
\newcommand{\rin}{\ensuremath{\mathit{R}_{\rm in}}}         
\newcommand{\lsn}{\ensuremath{\mathit{L}_{\rm SN}}}         
\newcommand{\tauross}{\ensuremath{\tau_{\mathrm{Ross}}}}                 


	\title{The diversity of supernovae and impostors shortly after explosion}

   \author{I. Boian
          \inst{1}
          \and
          J. H. Groh\inst{1}
          }

   \institute{Trinity College Dublin, The University of Dublin, College Green, Dublin, Ireland \\
              \email{boiani@tcd.ie} ; \email{jose.groh@tcd.ie}
             }

   \date{Received xxyyzz; accepted xxyyzz}

 
  \abstract
 {
Observational surveys are now able to detect an increasing number of transients, such as core-collapse supernovae (SN) and powerful non-terminal outbursts (SN impostors). Dedicated spectroscopic facilities can follow up these events shortly after detection. Here we investigate the properties of these explosions at early times. We use the radiative transfer code CMFGEN to build an extensive library of spectra simulating the interaction of supernovae and their progenitor's winds/circumstellar medium (CSM). We consider a range of progenitor mass-loss rates ($\mdot = 5 \times 10^{-4}$ to $10^{-2} ~\msunano$), abundances (solar, CNO-processed, and He-rich), and SN luminosities ($L = 1.9 \times 10^8$ to $2.5 \times 10^{10} ~\lsun$). The models simulate events $\simeq 1$ day after explosion, and we assume a fixed location of the shock front as $\rin=8.6 \times10^{13}$~cm. We show that the large range of massive star properties at the pre-SN stage causes a diversity of early-time interacting SN and impostors. We identify three main classes of early-time spectra consisting of relatively high-ionisation (e.g. \ion{He}{II} and \ion{O}{VI}), medium-ionisation (e.g. \ion{C}{III} and \ion{N}{III}), and low-ionisation lines (e.g. \ion{He}{I} and \ion{Fe}{II/III}). They are regulated by $L$ and the CSM density. Given a progenitor wind velocity $\vinf$, our models also place a lower limit of $\mdot \gtrsim 5 \times 10^{-4}\, (\vinf/150 \kms)\, \msunyr$ for detection of CSM interaction signatures in observed spectra. Early-time SN spectra should provide clear constraints on progenitors by measuring H, He, and CNO abundances if the progenitors come from single stars. The connections are less clear considering the effects of binary evolution. Nevertheless, our models provide a clear path for linking the final stages of massive stars to their post-explosion spectra at early times, and guiding future observational follow-up of transients with facilities such as the Zwicky Transient Facility.
}

   \keywords{supernovae: general --
                stars: massive stars --
                surveys --
                stars: mass-loss --
                stars: outflows --
                stars: evolution 
               }
   \titlerunning{Interacting SN and impostors shortly after explosion}
   \maketitle
%
\section{Introduction}
Most stars more massive than $\approx 8 ~\msun$ end their lives as supernovae (SNe). While the observed diversity in supernovae should reflect the variety of massive star progenitors, the exact links are not fully established.

There are various complementary methods to determine the progenitors of SNe. For a number of events, progenitors have been identified in pre-explosion images (\citealt{smartt15,vandyk17} and references therein). However, in general only limited photometric observations are available. SN lightcurves and spectra can give an indication of the ejecta mass, explosion energy and amount of synthesised $\ion{Ni}{}^{56}$, which can then be linked to the progenitor properties such as its initial mass \citep{woosley86}. For interacting events, lightcurves can be used to constrain some of the atmosphere/wind properties as well \citep[e.g.,][]{moriya14a}, which also map to a progenitor class. The environments of transients constrain the ages and metallicities of the nearby stellar populations and of the progenitor itself \citep{anderson15,xiao18}.

In addition to the techniques discussed above, early-time or ``flash'' spectroscopy provides substantial information on SN progenitors \citep{galyam14,groh14,smith15,grafener16,khazov16,yaron17}. Throughout their evolution, massive stars have high mass-loss rates ($\mdot \geq 10^{-6} \msunano$; \citealt{smith14araa}) that form a circumstellar medium (CSM). When the stars undergo an explosive event, the fast ejecta ($ v_{\mathrm{ej}} \approx 10000 ~\kms$) will crash into the slow CSM ($\vinf \approx 10 - 1000 ~\kms$) and the kinetic energy of the SN ejecta is converted to radiation. The radiation then photo-ionises the CSM, which becomes optically thick and the photosphere is formed in the CSM for a period of time ranging from days to a few years, depending on the CSM density and extension. Most optical emission lines such as H$\alpha$ are formed by recombination in a medium under photoionisation equilibrium. Therefore, a spectrum obtained at this stage should reflect the CSM density, velocity, and abundance profiles \citep{chevalier94,chugai01,dessart09,groh14,dessart15}.Using this technique, \cite{groh14} suggested that the progenitor of SN 2013cu is consistent with a luminous blue variable (LBV) or a yellow hypergiant (YHG; see also \citealt{grafener16}). Similarly \cite{shivvers15} found that the progenitor of SN 1998S could have been an LBV, YHG or massive YHG. \cite{yaron17} proposed a red supergiant (RSG) as the progenitor of SN 2013fs (see also \citealt{dessart17}). 

Due to the presence of a dense CSM, these SN show a 'wind breakout' \citep{moriya18} rather than a classical shock breakout. In this paper we cover a large range of mass-loss rates, which results in a large range of CSM densities. Therefore some of the SNe may show long-term interaction and if observed would be classified as type IIn SNe. However, others may show short-term interaction that may be missed by the observers if not detected early enough and would be classified as a type IIP, IIL, or IIb, based on the late-time spectra.

Since early-time spectroscopic observations and modelling of supernovae probe the most final stages of the star, they are a useful tool not only for linking supernovae to their progenitors but also in improving our knowledge of late-time massive star evolution. This is due to the fact that accurate observations of massive stars at late stages are impaired by their low numbers and sometimes dust-enshrouded surroundings, while modelling their evolution shows large dependencies on mass-loss, metallicities, rotational velocities \citep{ekstrom12,georgy13}, magnetic fields \citep{petit17}, and interaction with a companion \citep{pod92,langer12,eldridge17}.    
These factors play an important role in determining the ``explodability'' of a massive star \citep[e.g.,][]{renzo17,sukhbold18}, and the type and mass of the SN remnant \citep{fryer01,heger03,oconnor11} on which our understanding of gravitational wave observations rely \citep{abbott16,abbott17c,belczynski16}.

It is clear that there is a need for improved understanding of the final stages of massive star evolution and their links to SNe. In this article we present our grid of models that simulate the spectra of supernovae shortly after explosion ($ \approx 1$~day). Observations on a timescale of a day offer a good compromise between the time necessary to obtain transient spectra and the sensitivity to low-density CSM. We explore a wide range of progenitor mass-loss rates, surface abundances, and explosion luminosities to investigate the spectral diversity of SNe at early times. The set of models discussed here is part of an extensive library that will be made publicly available to the community. Our models complement the extensive observational efforts that have established the diversity of these events \citep[e.g.,][]{smith11,taddia13,khazov16}, and extend previous computational efforts \citep{dessart09,dessart17,groh14} to a large range of CSM abundances and densities. Given the rapid increase in the number of observed events with the advent of wide-field surveys such as the Zwicky Transient Facility (ZTF), our public library of models provides the community with a tool to estimate progenitor properties in a timely fashion. 

	
\section{CMFGEN radiative transfer modelling}
		We use the non-local thermodynamical equilibrium radiative transfer code CMFGEN \citep{hm98} to compute radiative transfer models of interacting SN and SN impostors at early stages after the explosion. In our setup, the code simulates the transport of radiative energy generated by the conversion of the kinetic energy of the supernova ejecta as it shocks against the progenitor's wind/CSM. The code requires a number of input physical parameters, such as the location of the inner boundary (\rin), luminosity of the event ($L$), and the progenitor's mass-loss rate \mdot, wind terminal velocity (\vinf) and surface abundances. We refer to \cite{groh14}, \cite{shivvers15}, and \citet{yaron17} for further details on the implementation. 
        
		We assume a steep density scale height with a gradient of $0.008$ \rin\ for the shocked region connected smoothly to an unclumped stationary wind which has a density profile $\rho \propto r^{-2}$ to represent the wind/CSM of the progenitor. We assume diffusion approximation at \rin\ and adjust the density to have a Rosseland optical depth $\tauross \gtrsim 100$ at \rin. We have ensured that the photosphere (where $\tauross =2/3$) is located in the progenitor wind for the radiation field and hydrodynamics of the inner region to minimally affect the spectra. We choose $\rin = 8.64 \times 10^{13}$ cm to correspond to the location of the shock front at $t = \simeq 1$ day after explosion, assuming a constant SN ejecta velocity of $10000$ \kms.
        
        For a given value of \rin\ and $L$ we can use the Stefan-Boltzmann law to define a flux temperature at the inner boundary:
           \begin{equation}
           \tin =  \Big( \frac{L}{4 \pi \rin^{2} \sigma } \Big) ^{\frac{1}{4}}
           \label{eq:sb}
           \end{equation}
         where $\sigma$ is the Stefan-Boltzmann constant. Therefore \tin\ is affected by both \rin\ and $L$ and a change in either parameter could modify the spectral morphology. In this paper we use a single \rin\ value for all models, both because we focus on a specific time after explosion (about 1 day) and to minimise the number of parameters that vary in our grid of models. We stress that additional models would be needed to explore the spectral morphology at times much larger than 1 day after explosion. 
         
        We compute models with $L = 1.9 \times 10^8$ to $2.5 \times 10^{10} ~\lsun$, $\mdot = 5 \times 10^{-4}$ to $10^{-2} ~\msunano$, $\vinf=150\,\kms$ and three sets of abundances (solar, CNO-processed, and He-rich). We also keep the same \vinf\ for all models in Figs. \ref{fig:spectra} and \ref{fig:luminosity} to minimise the number of parameters affecting the spectral morphology, and to avoid making a connection between quiescent wind parameters and pre-SN outflow properties. We discuss the effects of lower values of \vinf\ in Sect. 3. The models include \ion{Si}{}, \ion{P}{}, \ion{S}{} and \ion{Fe}{} at solar abundances. Table \ref{tbl:snprop} lists the model properties. 
             
        While CMFGEN can produce arbitrarily high resolution spectra, spectroscopic observations of SNe usually have low to medium resolution ($30 \leq \Delta v \leq 2000 ~\kms$), therefore Fig. \ref{fig:spectra} shows our spectra convolved with a Gaussian kernel of $\Delta v = 300 ~\kms$ (resolving power $R \sim 1000$). The high-resolution spectra can be found in Appendix \ref{app:resolutions} and in our online database. 
         
          Our models have several approximations to study the radiative transfer problem in interacting SNe. First, we do not solve for the time-dependent radiation hydrodynamics of the CSM. For a detailed discussion on these effects, see \citet{dessart17}. Our models are valid when the photosphere forms in the CSM, during which radiative equilibrium is still a reasonable approximation \citep{dessart09,groh14}. In addition, ignoring time-dependent effects allows the models to complete in a reasonable amount of time, which is crucial for producing a large grid of models that can fit the observations as is the goal of this paper. Second, we compute models with various $L$ and $\mdot$, although the time variation of $L$ should depend on the diffusion timescale, and thus $\mdot$. We note that the diffusion timescale is $\approx 10$ hours for $\mdot=10^{-2} ~\msunano$ and $\vinf=150~\kms$ (\citealt{chevalier11}), and that the same \mdot\ can lead to a variety of $L$ depending on the explosion energies \citep{moriya11}, asymmetry of the CSM or clumping/porosity in the CSM.  Thirdly, we assume a spherically-symmetric CSM, which may not be the case for some SNe (\citealt{smith17hsn}).
     
		\begin{table} 
		\caption{Physical properties of the CMFGEN radiative transfer simulations. The columns correspond to the SN $L$ in units of $10^9 ~\lsun$ $(L_\mathrm{9}$), progenitor's \mdot\ in units of $10^{-3} ~\msunano$ ($\mdot_\mathrm{3}$), flux temperature at $\tau=10$ (in kK), and the strongest optical spectral lines.  \label{tbl:snprop}}
		\begin{tabular}{c c c c}
		\hline
		\hline
		 $L_\mathrm{9}$ & $\mdot_\mathrm{3}$ & \tin & Spectral Lines \\
		\hline
		\hline
		\multicolumn{4}{c}{Solar abundances (H: $0.70$, He: $0.28$} \\
		\multicolumn{4}{c}{C: $3.02 \times 10^{-3}$, N: $1.18 \times 10^{-3}$, O: $9.63 \times 10^{-3}$)} \\
		\hline
	$0.19$&$0.5$&$19.2$ & \ion{H}{i} \\ 
		$0.19$ & $1.0$ &$19.1$& \ion{H}{i}, \ion{He}{i}\\ 
		$0.19$ & $3.0$ & $19.0$ & \ion{H}{i},\ion{He}{i}, [\ion{O}{iii}] \\ 
		$0.19$ & $10.0$ & $16.2$ &\ion{H}{i}, \ion{He}{i}, [\ion{O}{iii}],[\ion{N}{ii}] \\
        $0.39$ & $3.0$ & $22.6$ &\ion{H}{i}, \ion{He}{i}, \ion{C}{iii}, [\ion{O}{iii}] \\
        $0.78$ & $3.0$ & $26.9$ &\ion{H}{i}, \ion{He}{i/ii}, \ion{C}{iii/iv},  \ion{N}{iii}, [\ion{O}{iii}] \\
		$1.5$ & $0.5$ & $31.9$ & \ion{H}{i}, \ion{He}{ii} \\ 
		$1.5$ & $1.0$ & $32.2$ &\ion{H}{i},\ion{He}{ii},\ion{C}{iii/iv},\ion{N}{iii}\\ 
		$1.5$ & $3.0$ & $32.0$ &\ion{H}{i}, \ion{He}{ii},\ion{C}{iii/iv},\ion{N}{iii},[\ion{O}{iii}]\\ 
		$1.5$ & $10.0$ & $27.2$ &\ion{H}{i},\ion{He}{ii},\ion{C}{iii},\ion{C}{iv},\ion{N}{iii},\ion{Si}{iv},[\ion{O}{iii}]\\ 
        $3.1$ & $3.0$ & $38.0$ &\ion{H}{i}, \ion{He}{ii}, \ion{C}{iv}, \ion{O}{v}\\
        $6.3$ & $3.0$ & $45.3$ &\ion{H}{i}, \ion{He}{ii},  \ion{C}{iv}, \ion{N}{v}, \ion{O}{v} \\
		$25$ & $0.5$ & $64.4$ & - \\ 
		$25$ & $1.0$ & $64.4$ & \ion{H}{i}, \ion{O}{vi} \\ 
		$25$ & $3.0$ & $64.1$ & \ion{H}{i}, \ion{He}{ii},\ion{O}{vi}\\ 
		$25$ & $10.0$ & $55.9$ &\ion{H}{i}, \ion{He}{ii},\ion{O}{vi} \\
		\hline
		\multicolumn{4}{c}{CNO-processed models (H: $0.70$, He: $0.28$} \\
		\multicolumn{4}{c}{ C: $5.58 \times 10^{-5}$, N: $8.17 \times 10^{-3}$, O: $1.32 \times 10^{-4}$)}\\
		\hline
		$0.19$ & $0.5$ & $19.2$ & \ion{H}{i} \\ 
		$0.19$ & $1.0$ & $19.1$ & \ion{H}{i}, \ion{He}{i}\\ 
		$0.19$ & $3.0$ & $19.0$ & \ion{H}{i}, \ion{He}{i} \\ 
		$0.19$ & $10.0$ & $16.1$ & \ion{H}{i}, \ion{He}{i}, [\ion{O}{iii}] \\
         $0.39$ & $3.0$ & $22.6$ &\ion{H}{i}, \ion{He}{i}, \ion{N}{iii} \\
        $0.78$ & $3.0$ & $26.9$ &\ion{H}{i}, \ion{He}{i}, \ion{N}{iii}\\
		$1.5$ & $0.5$ & $31.9$ & \ion{H}{i}, \ion{He}{ii} \\ 
		$1.5$ & $1.0$ & $31.9$ & \ion{H}{i}, \ion{He}{ii}, \ion{N}{iii/iv}  \\ 
		$1.5$ & $3.0$ & $32.0$ &\ion{H}{i},  \ion{He}{ii}, \ion{N}{iii/iv} \\ 
		$1.5$ & $10.0$ & $27.3$ & \ion{H}{i} \ion{He}{ii}, \ion{N}{iii/iv} \\ 
        $3.1$ & $3.0$ & $38.0$ &\ion{H}{i}, \ion{He}{ii},  \ion{C}{iv}, \ion{N}{iv}\\
        $6.3$ & $3.0$ & $45.3$ &\ion{H}{i}, \ion{He}{ii}, \ion{N}{v} \\
		$25$ & $0.5$ & $64.4$ & - \\ 
		$25$ & $1.0$ & $64.4$ & \ion{H}{i}, \ion{He}{ii}, \ion{N}{v} \\ 
		$25$ & $3.0$ & $64.1$ & \ion{H}{i}, \ion{He}{ii}, \ion{N}{v}  \\ 
		$25$ & $10.0$ & $56.1$ & \ion{H}{i}, \ion{He}{ii}, \ion{N}{v}  \\
		\hline
		\multicolumn{4}{c}{\ion{He}{}-rich models (H: $0.18$, He: $0.80$} \\
		\multicolumn{4}{c}{ C: $5.58 \times 10^{-5}$, N:$8.17 \times 10^{-3}$, O:$1.312 \times 10^{-4}$)}\\
		\hline
		$0.19$ & $0.5$ & $19.2$ & \ion{H}{i}\\ 
		$0.19$ & $1.0$ & $19.2$ & \ion{H}{i}, \ion{He}{i}\\ 
		$0.19$ & $3.0$ & $19.1$ & \ion{H}{i}, \ion{He}{i}, [\ion{O}{iii}] \\ 
		$0.19$ & $10.0$ & $18.9$ & \ion{H}{i}, \ion{He}{i}, [\ion{O}{iii}] \\
        $0.39$ & $3.0$ & $22.7$ &\ion{H}{i}, \ion{He}{i}, \ion{N}{iii}, [\ion{O}{iii}] \\
        $0.78$ & $3.0$ & $26.9$ &\ion{H}{i}, \ion{He}{i}, \ion{N}{iii} \\
		$1.5$ & $0.5 $ & $32.0$ & \ion{H}{i} \\ 
		$1.5$ & $1.0$ & $31.9$ & \ion{H}{i}, \ion{He}{ii}, \ion{N}{iii/iv} \\ 
		$1.5$ & $3.0$ & $32.1$ & \ion{H}{i}, \ion{He}{ii}, \ion{N}{iii/iv}\\ 
		$1.5$ & $10.0$ & $31.7$ & \ion{H}{i}, \ion{He}{ii}, \ion{N}{iii/iv}\\ 
        $3.1$ & $3.0$ & $38.2$ &\ion{H}{i}, \ion{He}{ii}, \ion{N}{iv} \\
        $6.3$ & $3.0$ & $45.4$ &\ion{H}{i}, \ion{He}{ii}, \ion{N}{v}\\
		$25$ & $0.5$ & $64.7$ & - \\ 
		$25$ & $1.0$ & $64.5$ & \ion{He}{ii}, \ion{N}{v} \\ 
		$25$ & $3.0$ & $64.2$ & \ion{H}{i}, \ion{He}{ii}, \ion{N}{v}\\ 
		$25$ & $10.0$ & $63.8$ & \ion{H}{i}, \ion{He}{ii}, \ion{N}{v} \\
		\end{tabular}
		\end{table}
		
	\section{Diversity of interacting SNe and impostors}
    
     \begin{figure*}
     \center
 \includegraphics[width=0.87\textwidth]{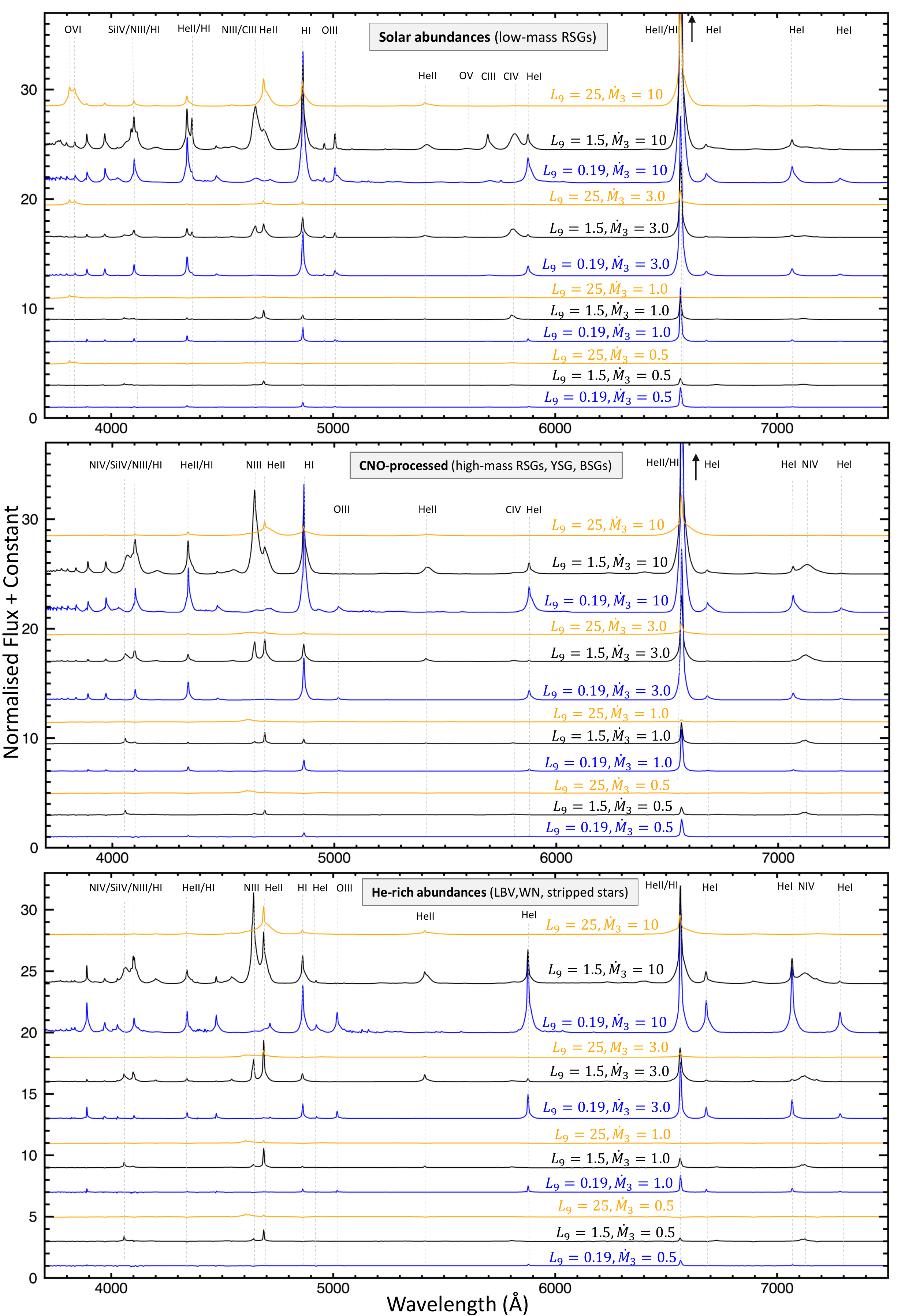}
 \caption{CMFGEN optical spectra of SNe interacting with their progenitor's wind at early times for solar (top panel), CNO-processed (middle), and He-rich abundances (bottom). The models have been convolved with a Gaussian kernel of $\Delta v = 300 ~\kms$ (resolving power $R \approx 1000$), continuum normalized, and shifted for clarity. The labels indicate the adopted values of $\mdot_{3}$ ($ 10^{-3} ~\msunano$) and $L_{9}$ ($10^9 ~\lsun$), and the strongest spectral lines are identified.}
 \label{fig:spectra}
 \end{figure*}

\begin{figure*}
     \center
 \includegraphics[width=0.87\textwidth]{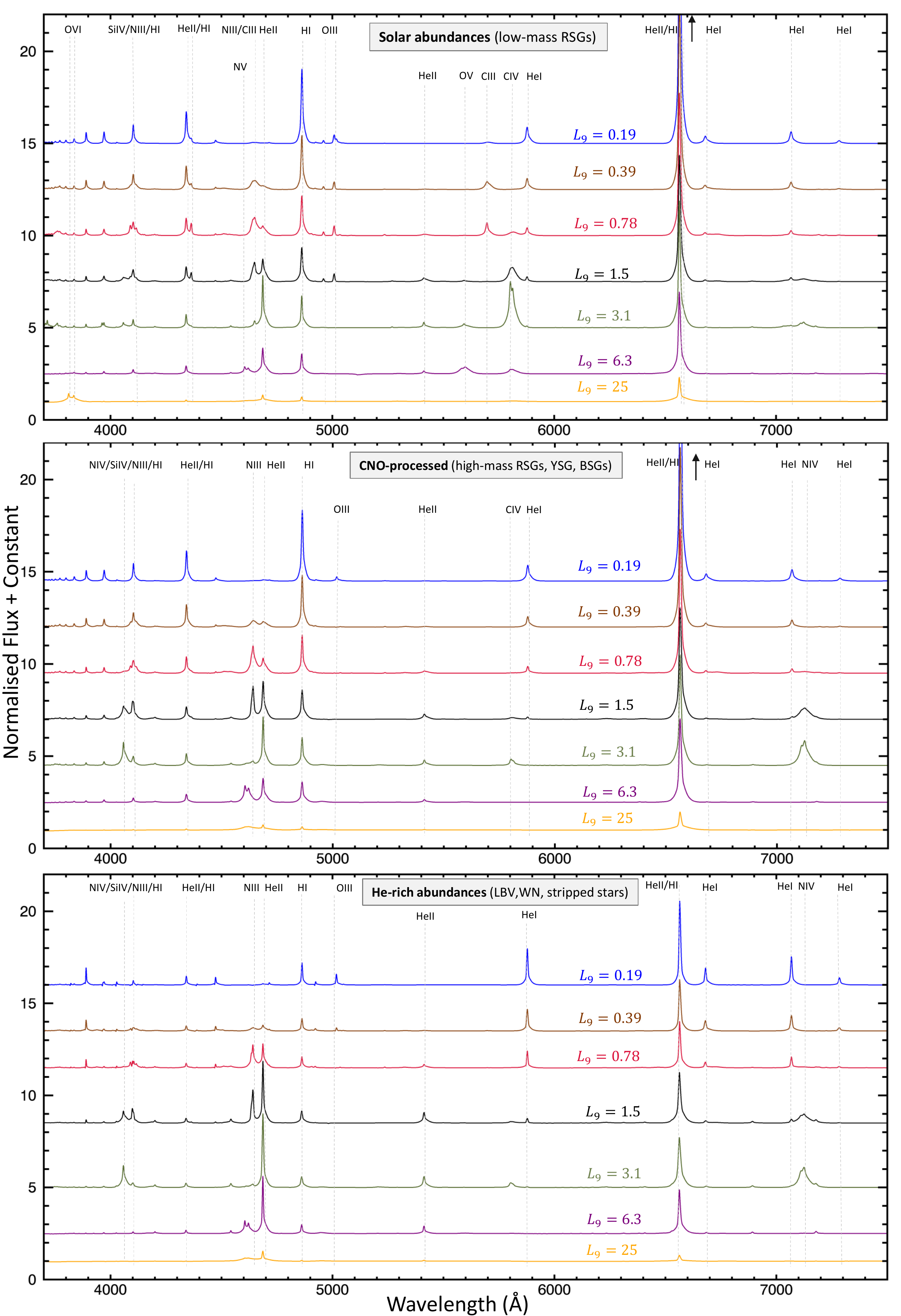}
 \caption{CMFGEN optical spectra of SNe interacting with their progenitor's wind at early times for solar (top panel), CNO-processed (middle), and He-rich abundances (bottom). The models have been convolved with a Gaussian kernel of $\Delta v = 300 ~\kms$ (resolving power $R \approx 1000$), continuum normalized, and shifted for clarity. All the models in this figure have $\mdot=3 \times 10^{-3} \msunyr$. The labels indicate the adopted values of $L_{9}$ ($ 10^9 ~\lsun$), and the strongest spectral lines are identified.}
 \label{fig:luminosity}
 \end{figure*}

	Our models show that the large range of physical parameters of massive stars before death produces a significant diversity of spectral morphologies in SNe . Broadly, we identify three main morphologies in transient optical spectra produced shortly after explosion in our solar abundance models at a medium resolution of $R=1000$ (Fig.~\ref{fig:spectra} \& ~\ref{fig:luminosity}, top panel).
    \begin{itemize}
    \item{High-ionisation spectra, characterized by \ion{O}{vi}, \ion{He}{ii}, \ion{N}{v}, and \ion{H}{i} emission lines. This morphology appears at high $L$ ($> 6.3 \times10^{9}~\lsun$) and resembles SN~2013fs;}
    \item{Medium-ionisation spectra, with strong \ion{He}{ii}, \ion{C}{iii}, \ion{N}{iii}, \ion{N}{iv}, \ion{C}{iv}, and \ion{He}{i} emission. These are seen at $L$ between $3.9 \times 10^{8} - 3.1 \times 10^{8} ~\lsun$ and might correspond to a solar-abundance version of SN 1998S observed at early times.}
    \item{Low-ionisation spectra, with \ion{H}{i}, \ion{He}{i}, and \ion{Fe}{ii} lines with P Cygni profiles. The models with high \mdot\ show [\ion{O}{iii}] and [\ion{N}{ii}]. This spectral morphology is present at relatively low $L$ ($\approx 1.9\times10^{8}~\lsun$) and resembles classical SN IIn, such as SN 1994W, as they would have been seen at early times.}
    \end{itemize}
	
    These morphologies arise from the dominant effect of $L$ on \tin, and thus on the ionisation structure of the CSM, and the strength of spectral lines from different ions. An increase in $L$ produces a more ionised CSM, and we see a shift towards lines of high ionisation stages. Perhaps surprisingly, the \ion{H}{i} emission decreases with increasing $L$, because the recombination rate coefficient decreases with the increase in the electron temperature ($T_{\mathrm{e}}$) that is seen in high $L$ models. Therefore, the $L = 1.9 \times 10^{8} ~\lsun$ models show the strongest \ion{H}{$\alpha$} emission for a given \mdot\ (Figs. \ref{fig:spectra} \& \ref{fig:luminosity}). 
    
     The spectral classes discussed above are meant to underline the effects of $L$ on the spectrum and guide the observers. However the changes on the spectrum are rather continuous, with cases falling in between the three described morphologies. One such example is the model with $L=6.3 \times 10^{9} ~\lsun$ which shows \ion{C}{iv} emission lines seen mostly in the medium ionisation case, but also shows \ion{N}{v} emission which belongs to the highly ionised spectral morphology (Fig. \ref{fig:luminosity}). Additionally, the abundances and the mass-loss rate will also play a role in which species are observable for each luminosity case.
    
    We find that changing the $\mdot$ of the progenitor affects the ionisation structure of the CSM as well, albeit to a lower degree than changing $L$ (we need $\mdot \gtrsim 0.1 \msunyr$ to generate a significant recombination of the CSM). For a given $L$, an increase in \mdot\ leads to stronger emission in all lines because of the increased density of the CSM. Also, since the density increases while the input of energy at the base of the wind is constant, the ionisation structure shifts towards less ionised ions (Fig. \ref{fig:spectra}). This is reflected in the ionisation stages seen in the spectra, e.g. higher \ion{N}{iii}/\ion{N}{iv} and \ion{C}{iii}/\ion{C}{iv} ratios for high values of \mdot. 
    
    The differences in the spectra due to changes in abundances are strong. In both CNO-processed and He-rich sets of models, we identify the same trend of spectral morphologies with respect to $L$ and \mdot\ discussed above. As expected, we find that the strength of spectral lines change monotonically according to the abundance of that element in the parameter range explored here. Our models with abundances corresponding to CNO-processed material show stronger N emission, weak C and no O lines (Fig. \ref{fig:spectra}, middle panel). The high-ionisation models show no \ion{O}{VI} even for high $L$. Medium-ionisation models develop strong \ion{N}{IV}, making them similar to SN 2013cu \citep{galyam14,groh14}. Low-ionisation models are not substantially affected in the models displayed here, although weak \ion{N}{II} lines could appear at certain combinations of $L$ and \mdot. The \ion{H}{$\alpha$} line is weaker in CNO-processed models compared to the corresponding solar abundance model. This is caused by the higher $T_{\mathrm{e}}$ of the CNO-processed models, which stems from the decrease of radiative losses in the CSM due to lower C and O abundances. The highest \mdot\ models shows the largest difference between the two different abundances. The decrease in emissivity due to the increase in $T_{\mathrm{e}}$ strongly affects the \ion{He}{ii} emission as well, and weakly the \ion{He}{i} emission. 
    
    The He-rich models show a substantial decrease in the ratio between \ion{H}{$\alpha$} and \ion{He}{ii} lines compared to the CNO-processed models (Fig. \ref{fig:spectra}, bottom panel). This is caused by the diminished H abundance. As for the CNO-processed models, almost no \ion{O}{} or \ion{C}{} emission are present in any of the He-rich models. Although the decrease in the \ion{H}{} abundance also lowers the opacity and raises $T_\mathrm{e}$, these changes do not substantially impact the CSM ionisation structure.

The \ion{H}{$\alpha$} and \ion{He}{ii} lines are the most degenerate with respect to \mdot\ and $L$. However, if both lines are detected in the spectrum, their abundances can be well constrained if the observations cover the optical range from 3700--7500~\ang, and either flux-calibrated spectra or multi-band photometry are available. As Fig.~\ref{fig:spectra} shows, this allows a determination of the ionisation structure of the wind, which breaks the degeneracy. E.g., while both a solar abundance, high $L$ model and a He-rich, low $L$ model can have similar \ion{H}{$\alpha$} strengths, the \ion{He}{ii} line will be much stronger for a higher He abundance. Analogously, \ion{He}{i} lines can be also employed.

The \ion{Fe}{} abundance is in general difficult to constrain from optical spectra alone. It could be determined from the \ion{Fe}{ii} lines present in the optical spectrum of some of the low-$L$ models. In contrast, the UV region is extremely populated by \ion{Fe}{} lines (Fig. \ref{fig:uv}) which provides a tool for determining the metallicities if UV data are available or for high-redshift SNe.

     \begin{figure*}
    \center
\includegraphics[width=0.99\textwidth]{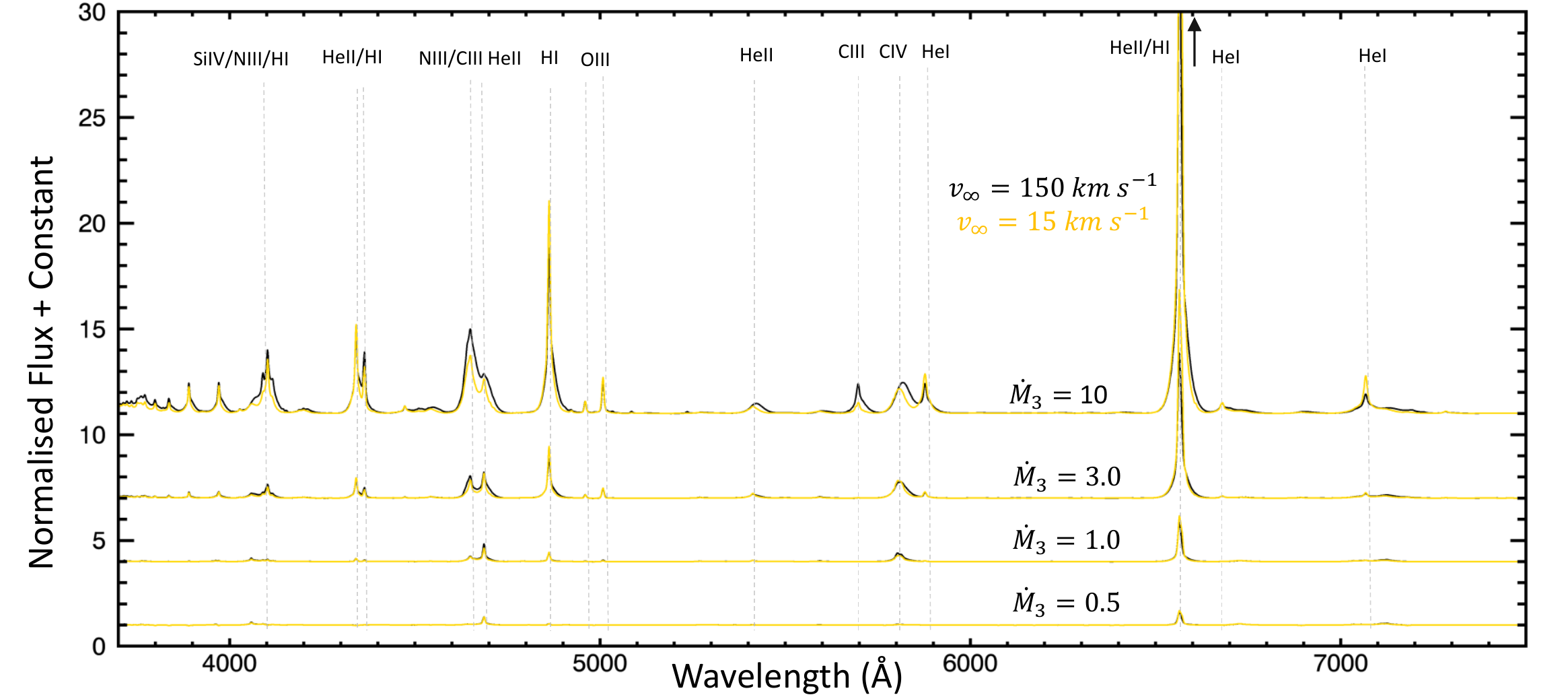}
\caption{CMFGEN optical spectra of SNe interacting with their progenitor's wind at early times for progenitors with $\vinf=150 ~\kms$ (models from Fig. \ref{fig:spectra}; black) and with $\vinf=15 ~\kms$ (overplotted on the corresponding high \vinf\ models; yellow). The models have been convolved with a Gaussian kernel of $\Delta v = 300 ~\kms$ (resolving power $R \approx 1000$), continuum normalized, and shifted for clarity. The labels indicate the adopted values of $\mdot_{3}$ ($ 10^{-3} ~\msunano$), and the strongest spectral lines are identified. All the models in this figure have solar abundances and $L=1.5 \times 10^{9} ~\lsun$.}
\label{fig:lowv}
\end{figure*}

Additional degeneracies appear due to a larger range of possible values of \vinf\ for events observed at low spectral resolution. This leads to an increased range of possible values of \mdot\ because of the mass continuity equation ($\rho \propto \mdot / \vinf$). Fig. \ref{fig:lowv} shows solar abundance models with $L=1.5 \times 10^9 ~\lsun$, $\vinf=15 ~\kms$, and resolution of $R = 1000$ together with the corresponding $\vinf=150~\kms$ models from Fig. \ref{fig:spectra}. Most of our models with the same density profile but different values of \vinf\ and \mdot\ are barely distinguishable at $R=1000$ (but see \citealt{grafener16}). Noticeable changes appear only for models with high CSM densities ($\mdot=10^{-2} ~\msunano$). In this case, a change in \vinf\ can result in an overlap of neighbouring spectral lines, which may lead to temperature changes observable in the \ion{N}{iii}/\ion{N}{iv} and \ion{C}{iii}/\ion{C}{iv} ratios.
 	 	
\section{Linking supernovae to their progenitors}\label{disc}

Early-time spectroscopy of SN offers three valuable clues on the nature of their progenitors: \mdot, \vinf, and abundances. 
However, theoretical models in recent years suggest a scenario in which the outflow properties and mass-loss mechanisms at the pre-SN stage \citep[e.g.,][and references therein]{fuller17} are markedly different than those seen in post-MS massive stars in the Galaxy and Magellanic Clouds \citep{smith17hsn}, which are expected to be at the He-core burning stage. Thus, linking \mdot\ and \vinf\ derived from early-time SNe to classes of massive stars characterized by those values may be misleading. This is highlighted by our lower limit of $\mdot \gtrsim 5\times 10^{-4} (150 ~\kms/\vinf) ~\msunyr$ for detection of clear spectroscopic evidence of SN-CSM interaction. This lower limit is higher than the typical \mdot\ of almost all massive stars in the local Universe, except for selected LBVs such as Eta Car \citep{ghm12} and others during S-Dor type outbursts \citep{smith14araa}.

For this reason, progenitor surface abundances may offer a more direct connection between pre- and post-SN spectra if abundances can be used as a proxy for the initial mass. This is generally the case for single star evolution \citep[e.g.,][]{gme13}. Binary evolution produces significantly more possible outcomes (see e.g. \citealt{eldridge17}) because of the large range of inferred initial orbital periods and mass ratios of massive binaries \citep{sana12,moe17}. Below we discuss some of the possible evolutionary connections of single and binary stars to our early-time spectral morphology, and the key spectroscopic diagnostics that could be used to constrain SN progenitors.

\begin{itemize}
    \item{Low-mass RSGs: they correspond to single star initial masses of 8--15~\msun, or less massive mass gainers in binary systems. These stars retain most of their massive \ion{H}{} envelope, with weak or no CNO processed material at the surface. These progenitors would give rise to early-time post-explosion spectra resembling our first set of models (Fig. \ref{fig:spectra} \& \ref{fig:luminosity} top panel). They should be identifiable using \ion{O}{vi} $\lambda \lambda 3811, 3834$ (high $L$), or \ion{C}{iii}$\lambda 5697$, \ion{C}{iv}$\lambda \lambda 5801,5811$ lines and \ion{N}{iii} $\lambda \lambda 4634, 4640$ (medium $L$), or \ion{C}{iii} $\lambda \lambda 4647,4650,4651$ (low $L$).}
     \item{Massive RSGs, YHGs, BSGs, LBVs, and WN stars: correspond to single star initial masses of 15--30~\msun. A range of initial periods should produce mass gainers, donors, and merger products that would be consistent with these classes of progenitors.  They exhibit CNO processed material at the surface as a result of rotational mixing, mass-loss, and/or binary interaction, with variable degree of H envelope stripping. They should show no \ion{O}{vi}$\lambda \lambda 3811, 3834$ at high $L$, possess \ion{N}{iv} $\lambda \lambda 4058, 7109, 7122$ and \ion{N}{iii}$\lambda \lambda 4634, 4640$ (medium $L$), and the \ion{C}{iii}$\lambda 5697$ and \ion{C}{iv}$\lambda \lambda 5801,5811$ lines should be absent. The ratio between \ion{He}{ii} and H lines should provide a reliable estimate of the H and He abundances, and thus on the nature of the progenitor. Higher He/H ratios favour late WN progenitors (these progenitors would show an enrichment of both \ion{N}{} and \ion{He}{} in their surface abundances and would produce post-explosion spectra similar to our third set of models shown in Fig. \ref{fig:spectra} \& \ref{fig:luminosity}, bottom panel), while mild He enrichment would be more consistent with RSGs and BSGs (their early-time post-explosion spectra would correspond to our second set of models in Fig.~\ref{fig:spectra} \& \ref{fig:luminosity}, middle panel). LBVs show intermediate He enrichment ($\sim0.65$ by mass).}
     
     We stress that the mass ranges and evolutionary scenarios above are model dependent and not comprehensive, in particular for binary evolution. In the single-star scenario, early-time SN spectra should provide clear constraints on progenitors by measuring H, He, and CNO abundances.

\end{itemize}

\subsection{Implications on Supernova impostors}
It is important to highlight that our models do not assume a terminal explosion as the source of energy at the base of the CSM. Because of that, they can also be applied to the class of SN impostors and intermediate transients. SN impostors are eruptive events that resemble low luminosity SNe, but are instead non-terminal explosions of massive stars \citep{vandyk00}.

It is well known that SN impostors show large spectral diversity, with some events exhibiting 'hot' spectra and others 'cool' spectra \citep{smith10}. While the models presented in this paper do not reach the lowest luminosities of SN impostors, they could be scaled down to reproduce SN impostors' spectra. To keep a similar \tin\ at low luminosities,  the inner radius would need to also be decreased according to Equation \ref{eq:sb}. To maintain the optical depth structure and line luminosities constant, \mdot\ would also have to be scaled using the relation $\mdot \propto L^{3/4} \times \vinf$ \citep[e.g.,][]{grafener16}. 

Some of our low \tin\ models do qualitatively resemble known 'hot' SN impostors such as SN 1997bs \citep{vandyk00}, SN 2000ch \citep{wagner04}, SN 2002bu \citep{smith11} or SN 2009ip \citep{smith10_sn2009ip}. When early-time observations of SN impostors become routinely available, our models should be able to reproduce the morphology and physical conditions of these events, allowing the determination of progenitor abundances and CSM properties. We will defer to future work an extension of the models to reproduce cool SN impostors.

\section{Observational challenges}\label{sec:survey}
\begin{figure*}
    \center
\includegraphics[width=0.87\textwidth]{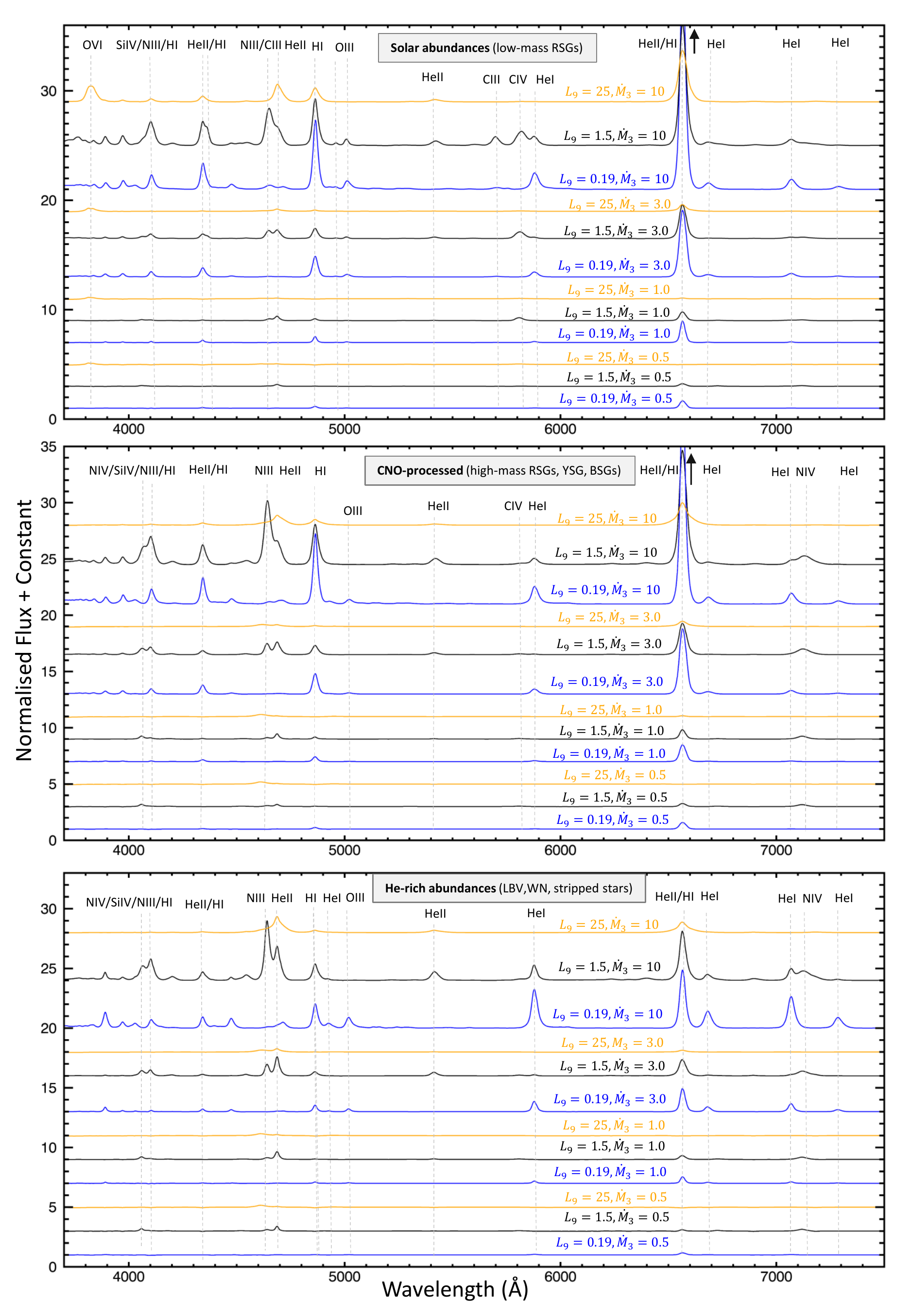}
\caption{Low-resolution ($\Delta v = 1000 ~\kms$, $R \approx 300$) CMFGEN optical spectra of SNe interacting with their progenitor's wind at early times for solar (top panel), CNO-processed (middle), and He-rich abundances (bottom). The models have been continuum normalized and shifted for clarity. The labels indicate the adopted values of $\mdot_{3}$ ($ 10^{-3} ~\msunano$) and $L_{9}$ ($ 10^9 ~\lsun$), and the strongest spectral lines are identified. }
\label{fig:lowres}
\end{figure*}

Our original CMFGEN models have arbitrarily high spectral resolution. However, spectroscopic observations of transients typically have low to medium resolution due to the need for fast spectroscopic follow-up of faint events. This is a continuing trend even with new surveys such as ZTF \citep{bellm17}, which will employ the Double Spectrograph on the Palomar 200'' (resolving power of $R \approx 1000-10000$), or the Spectral Enegry Distribution Machine ($R \approx 100$, \citealt{blagorodnova18}) on the Palomar 60'' telescope. For this reason, we have convolved our spectra with a Gaussian kernel of $\Delta v =1000 ~\kms$, i.e. resolving power $R \approx 300$, to account for the lower end of spectrograph resolutions. Fig. \ref{fig:lowres} shows the low-resolution spectra for the models in Fig. \ref{fig:spectra}. It can be easily seen by comparing figures \ref{fig:spectra} and \ref{fig:lowres} that the main trends identified in the higher resolution spectra and described in the text can still be observable at lower resolution. However, when fitting observed events, the errors will increase as differences between spectra will be harder to disentangle. This can be problematic specially for events with spectra characterized by medium-ionization lines.

\section{\bf Summary and Conclusions}

In this paper we have presented a systematic investigation of spectra that simulate SN interacting with their progenitor's wind $\approx 1$ day after explosion. We have covered a large range of SN luminosities ($L= 1.9 \times 10^{8} - 2.5 \times 10^{10} ~\lsun$), progenitor mass-loss rates ($\mdot = 5 \times 10^{-4} - 10^{-2} ~\msunano$), and progenitor's surface abundance scenarios (solar, CNO-processed, and He-rich).
We summarise our main results below.
\begin{enumerate}
\item For the luminosity range considered here the early-time spectra can be classified in three main categories based on ionisation levels.  Low-ionisation spectra show mainly \ion{H}{i} and \ion{He}{i} lines. In the medium-ionisation case \ion{C}{iii} or \ion{N}{iii} lines are present, while high-ionisation spectra contain \ion{He}{ii}, \ion{N}{v} or \ion{O}{v/vi} lines. 

\item Our models show that progenitors with mass-loss rates lower than $\mdot = 5 \times 10^{-4}\, (\vinf/150 \kms) ~\msunano$ will not show detectable interaction signatures in the post-explosion spectra.

\item Because of the unknown nature of pre-explosion mass loss, it is challenging to connect CSM densities to different progenitor types. On the other hand, changes in the surface abundances of the progenitors have observable effects on the early-time post-explosion spectra. Early-time SN spectra thus provide constraints on progenitors by measuring H, He, and CNO abundances if the progenitors come from single stars. The connections are less clear considering the effects of binary evolution.

\item The effects discussed above are also observable in low resolution spectra, hence they could be used for linking SN to their progenitors. However, low resolution spectra may yield higher errors in the parameter determinations due to the poorer constraints on the CSM velocity and progenitor's mass-loss rate.

\item The spectra presented in this paper can be used to characterise not only events with properties in the parameter range presented here, but they can be adapted to fit lower luminosity events, such as SN impostors, or events observed at different epochs after explosion. 
\end{enumerate}

Our models provide a clear path for linking the final stages of massive stars to their post-explosion spectra at early times, and guiding future observational follow-up of transients with facilities such as ZTF.

\begin{acknowledgements}
I.B. acknowledges funding from a Trinity College Postgraduate Award through the School of Physics, and J.H.G. acknowledges support from an Irish Research Council New Foundations Award 206086.14414 "Physics of Supernovae and Stars". 
\end{acknowledgements}

\bibliographystyle{aa} 

\bibliography{refs} 

\begin{appendix}
\section{Supplemental Figures}
\label{app:resolutions}
\begin{figure*}[h!]
\includegraphics[width=0.87\textwidth]{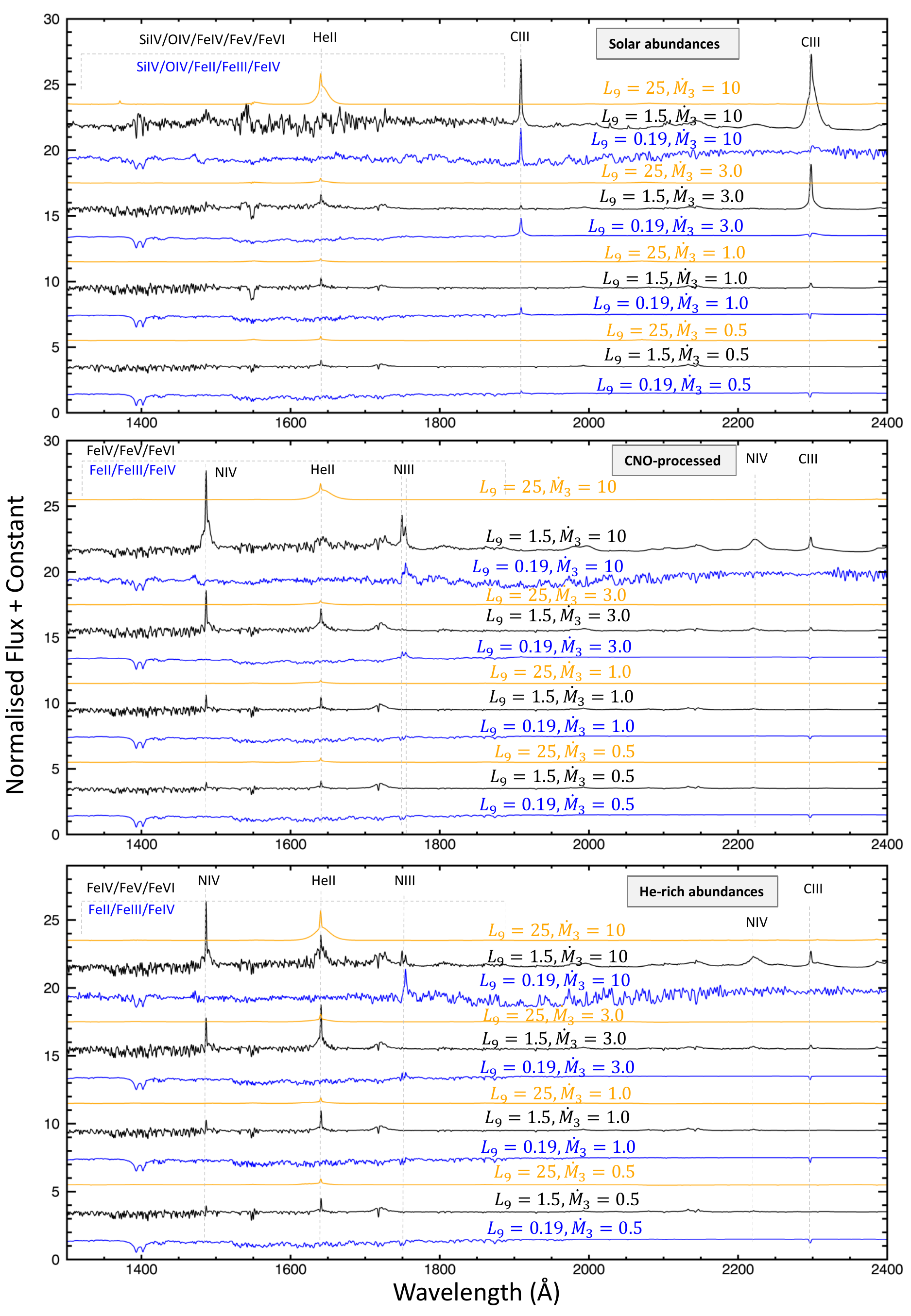}
\center
\caption{Ultraviolet spectral region of the CMFGEN models shown Fig. \ref{fig:spectra}. The models have been continuum normalized and shifted. The labels indicate the adopted values of $\mdot_{3}$ ($10^{-3} ~\msunano$) and $L_{9}$ ($10^9 ~\lsun$), and the strongest spectral lines are identified.}
\label{fig:uv}
\end{figure*}
\begin{figure*}
\includegraphics[width=0.87\textwidth]{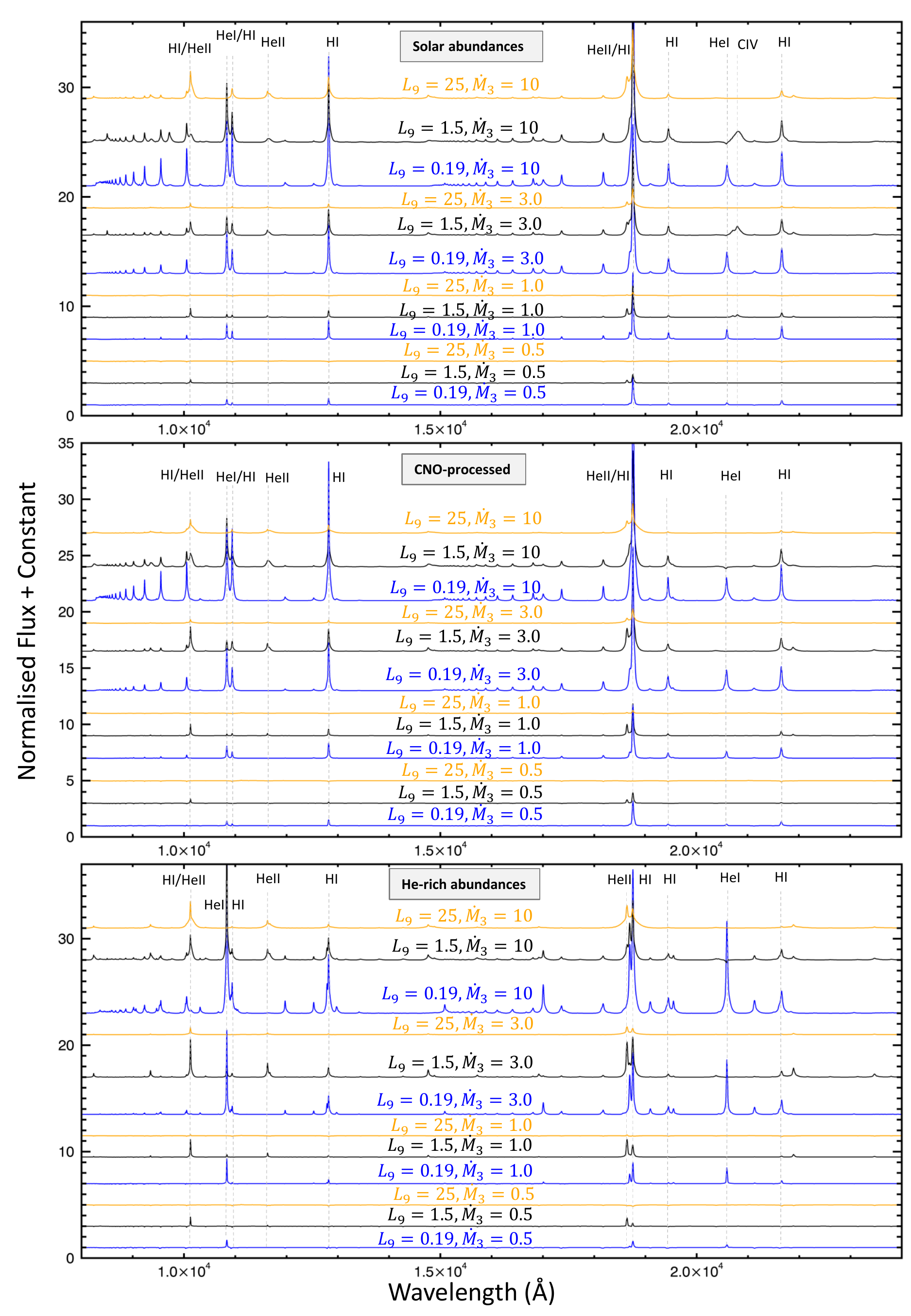}
\center
\caption{Similar to Fig. \ref{fig:uv}, but for near-infrared spectral region.}
\label{fig:nir}
\end{figure*}

 \begin{figure*}
 \includegraphics[width=0.87\textwidth]{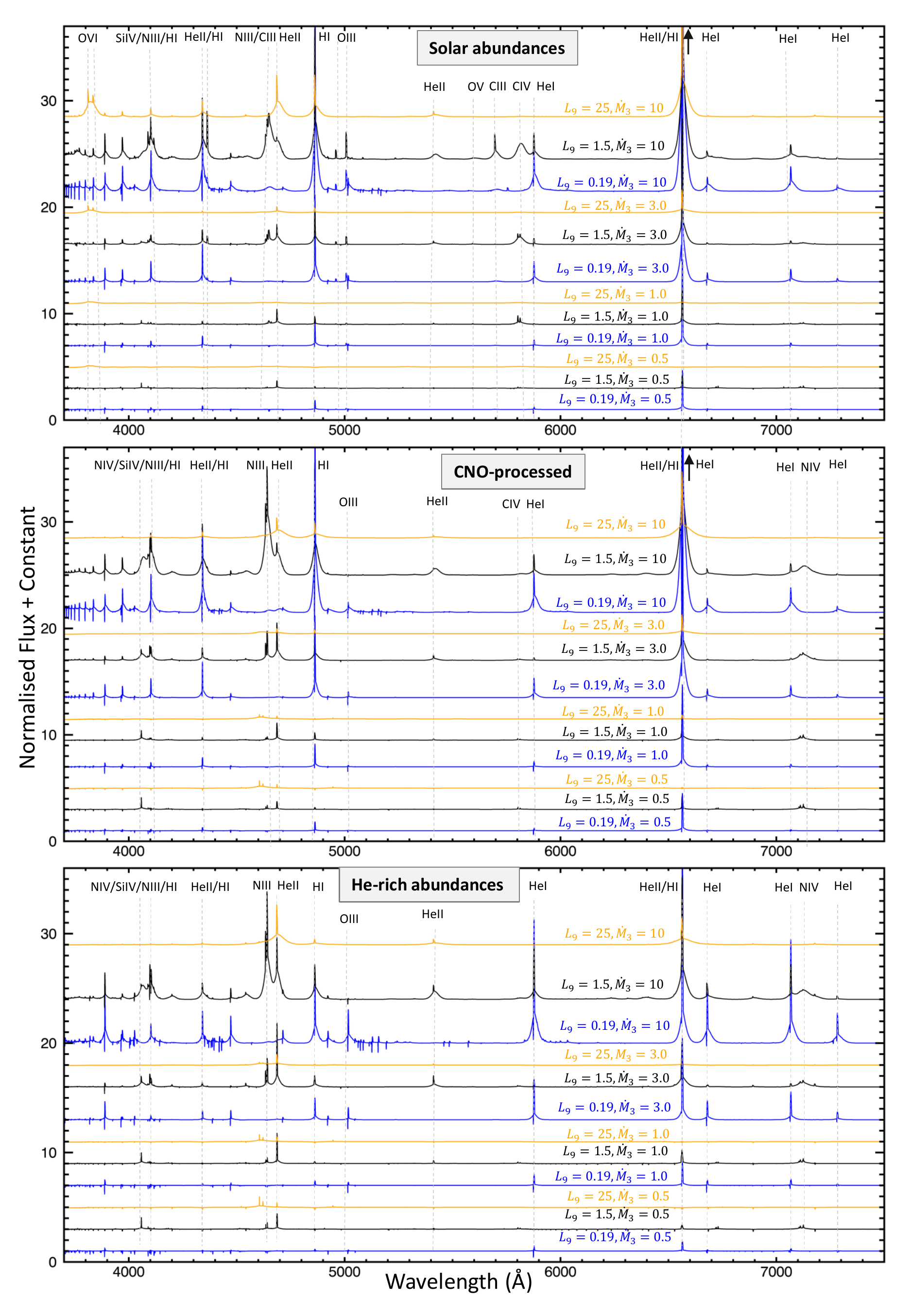}
 \center
 \caption{High-resolution CMFGEN optical spectra of SNe interacting with their progenitor's wind at early times. The models have been continuum normalized and shifted. The labels indicate the adopted values of $\mdot_{3}$ ($ 10^{-3} ~\msunano$) and $L_{9}$ ($ 10^9 ~\lsun$), and the strongest spectral lines are identified.}
 \label{fig:high}
 \end{figure*}

\end{appendix}


\end{document}